\documentclass[aps,prb,twocolumn,floatfix,nofootinbib]{revtex4}
\usepackage{graphicx,epsfig,array}
\usepackage[toc]{appendix}
\usepackage{color,graphicx,amssymb,amsmath}
\usepackage[utf8]{inputenc}
\usepackage{tensor,upgreek}
\usepackage{hyperref}

\begin{document}
	\title{Memory efficient Fock-space recursion scheme for computing many-fermion resolvents.}
	\author{Prabhakar}
		\affiliation{School of Physical Sciences, National Institute of Science Education and Research, a CI of Homi Bhabha National Institute, Jatni 752050, India}
		\author{Anamitra Mukherjee$^*$} 
	\affiliation{School of Physical Sciences, National Institute of Science Education and Research, a CI of Homi Bhabha National Institute, Jatni 752050, India}
	\date{\today}
	
	\begin{abstract}
	A fundamental roadblock to the exact numerical solution of many-fermion problems is the exponential growth of the Hilbert space with system size. It manifests as extreme dynamical memory and computation-time requirements for simulating many-fermion processes. 
		Here we construct a novel reorganization of the Hilbert space to establish that the exponential growth of dynamical-memory requirement is suppressed inversely with system size in our approach. Consequently, the state-of-the-art resolvent computation can be performed with substantially less memory. The memory-efficiency does not rely on Hamiltonian symmetries, sparseness, or boundary conditions and requires no additional memory to handle long-range density-density interaction and hopping. We provide examples calculations of interacting fermion ground state energy, the many-fermion density of states and few-body excitations in interacting ground states in one and two dimensions. 	
		\end{abstract}

	\maketitle
	\section{Introduction}
	\label{sec:intro}	
The exploration of complex quantum systems relies on studying many-fermion correlations. For example, the many-fermion density of states can be used to study many-body localization in both fermionic \cite{mbl-logan-prl} and spin \cite{mbl-huse,mbl-excited} systems. Stability of few-fermion bound complexes \cite{few-ed-1}, the passage from few to many body systems \cite{few-many}, chaos in few-body system \cite{few-chaos} and the existence of topological edge modes in interacting systems \cite{int-top} require the knowledge of the full many-fermion spectrum. In the study of few-body excitations such as photoemission spectroscopy \cite{arpes} and dynamical susceptibility \cite{Dyn_spin_susc_RMP,Dyn_S}, the Lanczos scheme \cite{lanczos-dagotto,cheby} and variants such as the shift-inverted technique \cite{pivot-optimal,shift-invert} are advantageous. Similarly, the density matrix renormalization group (DMRG) is extremely useful for obtaining ground states of quasi-one-dimensional systems \cite{dmrg-rmp}. However, for the former class of problems requiring full many-fermion density of states, there is currently no viable alternative to exact diagonalization. The well-known drawback of exact-diagonalization (ED) is the exponential growth of Hilbert space with system size. This paper presents a memory-efficient approach to compute many-fermion resolvent, from which the many-body density of states can be extracted. As an added benefit, it also allows the calculation of few-fermion excitations in interacting ground states. 
{\let\thefootnote\relax\footnote{{$^*$Corresponding author: anamitra@niser.ac.in}}}

For a $\mathcal{N}$-fermion lattice system, the many-fermion correlation functions are the vacuum expectation of the time-ordered product of $\mathcal{N}$ creation (annihilation) operators, all acting simultaneously at $t$ ($t^\prime$) with $t^\prime >t$. Its Fourier transform defines the many-fermion resolvent, which contains a plethora of information about the many-fermion system, such as the many-fermion density of states (DOS) and ground state energy. The computation of the resolvent amounts to brute-force inverting or solving a set of equations for a matrix of the dimension of the Hilbert space whose size grows exponentially with the number of lattice sites. Hence, suppressing the memory requirement of the many-fermion resolvent is highly desirable.

Here, we develop a memory-efficient computational technique for the exact calculation of many-fermion resolvent, also referred to as the many-fermion Green's functions. We establish that the scheme suppresses the exponential growth of the memory (RAM) requirement \textit{inverse in system size}.
For a $\mathcal{N}$ spinless fermions on $\mathcal{L}$ sites, the $O(1/\mathcal{L})$ scaling of memory, reduces RAM requirement of state-of-the-art calculation of 508Gb \cite{ravi-mbl} to 160Gb at half-filling ($\mathcal{N}=10$ for $\mathcal{L}=20$) and can allow access to $\mathcal{L}=22$ with 1.3Tb RAM instead of 7.5Tb for direct-inversion (DI). The method is independent of matrix sparsity and boundary conditions and can handle long-range hopping or long-range density-density interaction at no additional memory overhead. 

Section 2 presents the necessary definitions for setting up the problem. In section 3, we introduce the notion of a lattice in the Fock-space constructed from the many-fermion Hilbert space, discuss its structure and present a recursion algorithm to and detail compute the many-fermion correlation function. In section 4, we define a model Hamiltonian for interacting spinless fermions and show ground-state energies benchmarked against exact diagonalization, the many-fermion density of states, and two-hole spectra in fill and partially filled ground states in one and two dimensions. We conclude the paper in section 5.

	\section{Setup of the problem}
	\label{sec: setup}
	As mentioned above, we consider $\mathcal{N}$ spinless fermions on a $\mathcal{L}$ site chain. The dimension of the Hilbert space is $^\mathcal{L}C_\mathcal{N}$.
	We start by defining the many-fermion eigenvalue problem for $\mathcal{N}$ spinless-fermions on a $\mathcal{L}$ site chain as $\hat{H}|\lambda^\mathcal{N}\rangle=E^{\mathcal{N}}_{\lambda}|\lambda^{\mathcal
		{N}}\rangle$, with $\{E^{\mathcal{N}}_\lambda\}$ and $\{|\lambda^\mathcal{N}\rangle\}$ denoting the $\mathcal{N}$-fermion eigenvalues and eigenvectors, respectively. Among them  $|\psi_{0}^{\mathcal{N}} \rangle$ and
	$E^{\mathcal{N}}_0$ are the $\mathcal{N}$-fermion ground-state vector and
	ground-state energy respectively. In what follows, we set $\hbar=1$. 
	We then consider the $\mathcal{N}$-fermion retarded Green's function evaluated in the $\mathcal{N}$-fermion vacuum state $|0\rangle$. Starting from the $\mathcal{N}$ fermions created at time $t$ in $|0\rangle$ and destroyed at $t^\prime$ with $t^\prime >t$ and using standard Fourier transformation, we obtain the frequency space retarded $\mathcal{N}$-fermion Green's function, $\mathcal{G}^{\mathcal{N}}_{a_1,...,a_\mathcal{N};a_1^\prime,...,a_\mathcal{N}^\prime}(\omega)$ as $\langle
	0|c_{a_\mathcal{N}}...c_{a_1}\hat{\mathcal{G}}c^{\dagger}_{a_1^\prime}..
	.c^{\dagger}_{a_\mathcal{N}^\prime}|0\rangle$, with $\hat{\mathcal{G}}(\omega)\equiv
	(\omega-\hat{H}+i\eta)^{-1}$. The set of 2$\mathcal{N}$ subscripts (primed and unprimed), denote fermion positions.
	The full $\mathcal{N}$-fermion 
	Hilbert space is generated by considering all permutation of the fermion
	positions, giving the Hilbert space of dimension $^\mathcal{L}C_\mathcal{N}$. For brevity
	of notation we denote these $\mathcal{N}$-fermion Hilbert space basis vectors by
	lowercase italicized Latin letters with a subscript $\mathcal{N}$ denoting the number of fermions in the state. Thus, $\mathcal{G}^{\mathcal{N}}_{a_1,...,a_\mathcal{N};a_1^\prime,...,a
		_\mathcal{N}^\prime}(\omega)\equiv \mathcal{G}^{\mathcal{N}}_{j_{\mathcal{N}};j_{\mathcal{N}}^\prime}(\omega)$.  
	In the Lehmann representation, the $\mathcal{N}$-fermion retarded Green's function has the following form:
	\begin{equation}
		\mathcal{G}^{\mathcal{N}}_{j_{\mathcal{N}};j_{\mathcal{N}}^\prime}(\omega)=\sum_{\lambda^\mathcal{N}}\frac
		{\langle
			j_{\mathcal{N}}|\lambda^\mathcal{N}\rangle\langle\lambda^\mathcal{N}|j_{\mathcal{N}}^\prime\rangle}{\omega-E
			^\mathcal{N}_{\lambda}+i\eta}
		\label{eqn1}
	\end{equation}
	We have set the energy of the vacuum state to be zero, and $\eta$ is the usual regulator. The $\mathcal{N}$-fermion Green's function matrix, denoted by square brackets $[\mathcal{G}^{\mathcal{N}}(\omega)]$ is generated by running over all $\mathcal{N}$-fermion basis elements. The $\mathcal{N}$-fermion spectral function matrix in terms of the imaginary part of $[\mathcal{G}^{\mathcal{N}}(\omega)]$ is given by $[\mathcal{D}^{\mathcal{N}}(\omega)]\equiv -1/\pi
	Im\{[\mathcal{G}^{\mathcal{N}}(\omega)\}$. Finally the well-known $\mathcal{N}$-fermion spectral function $A^{\mathcal{N}}(\omega)$ is defined as the trace of $[\mathcal{D}^{(\mathcal{N})}(\omega)]$ over the many-fermion basis states. 
	The $\mathcal{N}$-fermion ground-state energy is extracted by determining the location of the \textit{lowest energy peak} of $A^{\mathcal{N}}(\omega)$. We also mention that $(\mathcal{N}-2)$-fermion spectral function matrix $[D^{\mathcal{N}-2}(\omega)]$ and $A^{\mathcal{N}-2}(\omega)$ are similarly extracted from ($\mathcal{N}-2$)-fermion Green's function matrix $[\mathcal{G}^{\mathcal{N}-2}(\omega)]$. 
	
	Thus the task at hand is computing $[\mathcal{G^N}(\omega)]$. This  involves inverting a  ($^\mathcal{L}C_\mathcal{N}$)-dimensional matrix that grows exponentially with $\mathcal{L}$. We provide a highly memory-efficient scheme for such inversions in the next section.

	\section{Fock-space Recursive Green's function}
	\label{sec:method}
	\begin{figure*}[h]
		\centering{
			\includegraphics[width=18cm, height=13.6cm, clip=true]{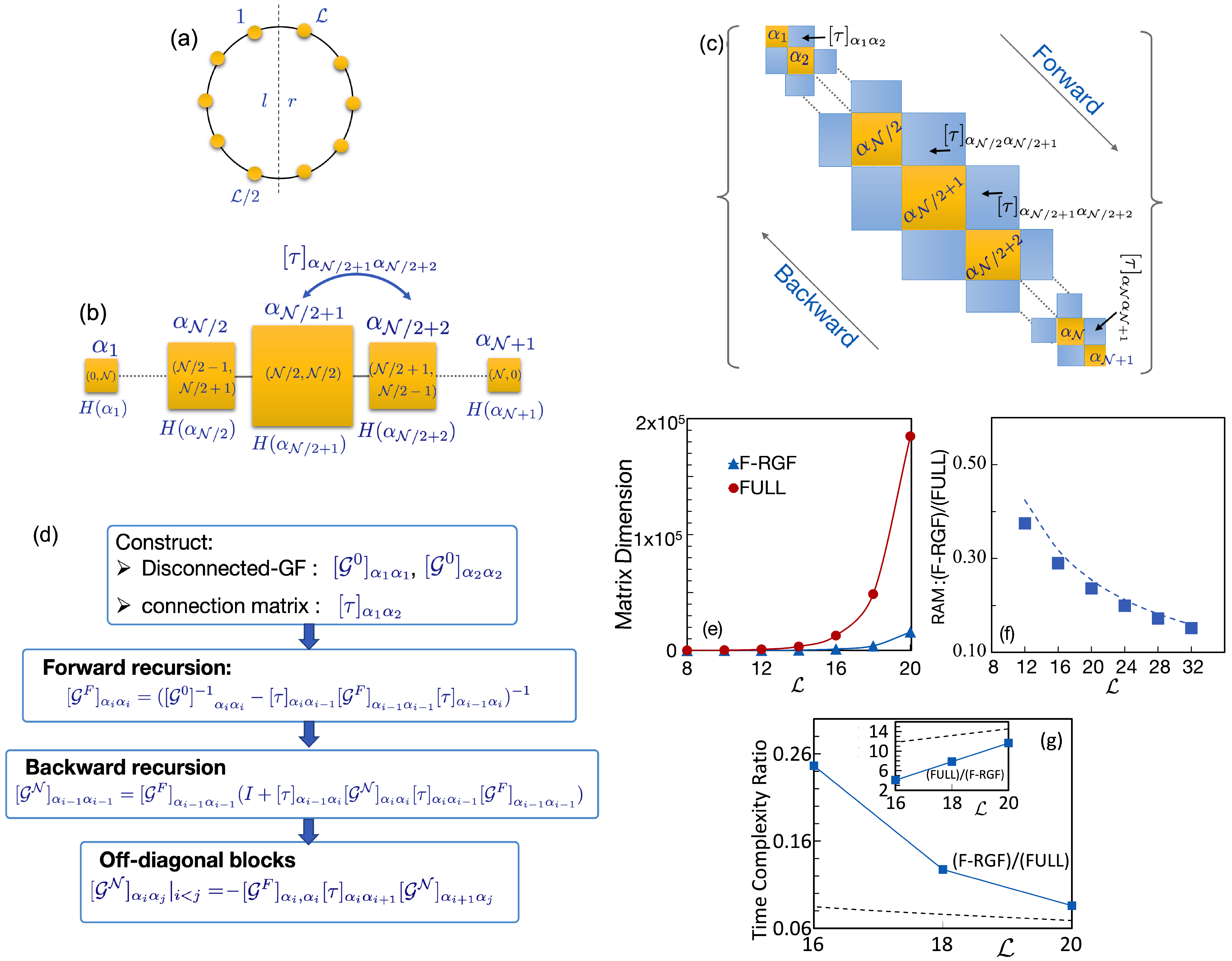}} 
		\caption{
			\textbf{Fock-space lattice, recursive algorithm \& memory efficiency and computation time advantage.} (a) shows a one-dimensional $\mathcal{L}$ sites lattice partitioned by a dashed line. The scheme also works for equal lattice bipartition with open boundaries. (b) Schematic of matrix-valued Fock-space lattice with nearest neighbor (nn) hopping connectivity. Squares with $\alpha_i$ labels denote left (right)-half occupations $n_l=i-1$ ($n_r=\mathcal{N}-{i+1}$), and   $H({\alpha_{i}})$ labels the Hamiltonian of $\alpha_{i}^{th}$ sector. The nn connection matrices $[\tau]_{\alpha_{i},\alpha_{i\pm1}}$ facilitate fermion number fluctuations between sectors, as shown between $\alpha_{\mathcal{N}/2+1}$ and $\alpha_{\mathcal{N}/2+2}$. (c) shows the block-tridiagonal structure of $H$. (d) provides the Fock-space recursive Green's function flowchart.
			Here, $[\mathcal{G}^0]_{\alpha_{i}\alpha_{i}}$, $[\mathcal{G}^{F}]_{\alpha_i,\alpha_i}$ and $[\mathcal{G}^{\mathcal{N}}]_{\alpha_i,\alpha_i}$ are the `disconnected, `forward-connected' and `full' Green's function $\alpha_i^{th}$ sector respectively.  
			$[\mathcal{G}^{\mathcal{N}}] _{\alpha_i,\alpha_j}$ with $(i\neq j)$ denote off-diagonal full Green's function blocks.  
			(e) shows the maximum matrix dimension for brute force direct inversion \& F-RGF at half-filling. 
			(f) Ratio of maximum RAM requirement in F-RGF that stores \textit{two} matrices of $\alpha_{\mathcal{N}/2+1}$  sector dimension to direct inversion at half-filling (symbols). 
			Exact theoretical scaling of $16/\pi\mathcal{L}$ (dashed line) is obtained from computed the ratio analytically and establishes $O(1/\mathcal{L})$ memory scaling.  (g) The main panel shows the computation time ratio for calculating many-fermion Green's function using F-RGF to ED or direct inversion (DI) as a function of $\mathcal{L}$ at half filling. The inset shows the inverse of the time ratio in the main panel. The dashed lines are the theoretical scaling, while the symbols denote code execution times in the main panel and inset. We use the ZHEEV routine for ED calculation on Intel E5-2683v4 2.10GHz processors. The details are discussed in the text.}
		\vspace{-0.0cm}
		\label{f-1}
	\end{figure*}

	The technique rests on two steps, first we present a construction of Fock-space sectors by suitably grouping many-fermion basis states. This leads to a matrix-valued lattice in the Fock-space and allows a `block tri-diagonal' representation of the Hamiltonian matrix. The second step generalizes the well-known recursive Green's function method \cite{rgf-old-1,block-inversion,rgf-der, Thouless_1981,rgf-formalism, parallel-rgf} to this Fock-space lattice to obtain $[\mathcal{G^N}(\omega)]$. We refer to the scheme as \textit{Fock-space Recursive Green's Function} (F-RGF).
	
	\textit{i. Fock-space lattice:}
	We partition the $\mathcal{L}$-site lattice into two halves as in Fig.~\ref{f-1} (a) and label the $\mathcal{N}$-fermion basis states by $|n_ln_r\rangle$, the fermion occupation on the left ($n_l$) and right ($n_r=\mathcal{N}-n_l$) halves. Under hopping of any range and open/closed boundary conditions, $|n_ln_r\rangle$ maps either to $|n_l\pm 1,n_r\mp 1\rangle$, implying fermion hopping across the geometric divide, or back to itself accounting for all hopping processes that conserve $n_l$ and $n_r$. Thus, the Hilbert space is decomposed into a direct sum of \textit{Fock-space sectors} 
	and connecting hopping matrices. The Fock-space sectors contain all non-local repulsion terms. 
	In Fig.~\ref{f-1} (b), $H$ is represented as a Fock-space lattice with $\mathcal{N}+1$ Fock-space sectors, $\alpha_{n_l+1}$ labelling the sector $(n_l, n_r)$ and, nearest-neighbor(nn) sector-hopping matrices $[\tau]_{\alpha_i,\alpha_j}$. 
%
Fig.~\ref{f-1} (c) depicts the block-tridiagonal form of $H$ that ensures the same structure for $(\omega+i\eta)I-H$, which is the inverse of $[\mathcal{G^N}(\omega)]$.

	\textit{ii. Recursion scheme :} The block-tridiagonal form substitutes $(^\mathcal{L}C_\mathcal{N})$-dimensional inversion by multiple smaller matrix inversions and a recursion scheme to obtain $[\mathcal{G^N}(\omega)]$. In this sub-section, the $\omega$ arguments of the Green's function are suppressed for clarity.
	
	The algorithm to obtain retarded Green's function $[\mathcal{G}^{\mathcal{N}}(\omega)]$,
	consists of a forward and backward recursion. We refer the reader to literature\cite{rgf-der, rgf-old-1} for detailed derivation of this well-established formalism, derived for non-interacting (one-body) Green's functions. Here we apply the same formalism on the one-dimensional Fock-space lattice that allows us to compute the many-fermion Green's function. The method involves `forward' and 'disconnected' Green's functions $[\mathcal{G}^{0}]_{\alpha_i\alpha_i}$ and $[\mathcal{G}^{F}]_{\alpha_i\alpha_i}$ respectively, and the connection matrices $[\uptau]_{\alpha_i,\alpha_j}$.  $[\mathcal{G}^{0}]_{\alpha_i\alpha_i}$ is the retarded Green's function for the $\alpha_i$ sector  defined as $(\omega+i\eta-H(\alpha_i))^{-1}$, with the corresponding sector Hamiltonian $H(\alpha_i)$. It is called `disconnected' as this Green's function does not involve the nn sector hopping matrices. The forward Green's function for sector $\alpha_i$ is defined in Eq.~\ref{eqn2}. The recursion algorithm is schematically shown in Fig.~\ref{f-1} (d).

	(a) The \textit{forward connected} Green's function can be calculated by the following recursive equation:
	\begin{align}
		[\mathcal{G}^{F}]_{\alpha_i\alpha_i}^{-1}=[\mathcal{G}^{0}]^{-1}_{\alpha_i\alpha_i}-[\uptau]_{\alpha_i\alpha_{i-1}}[\mathcal{G}^{F}]_{\alpha_{i-1}\alpha_{i-1}}[\uptau]_{\alpha_{i-1}\alpha_i}
		\label{eqn2}
	\end{align}
	We start from the leftmost sector labelled by $\alpha_1$ of the Fock-space lattice. Since there are no sectors to its left, from the above equation, $[\mathcal{G}^{F}]_{\alpha_i\alpha_i}=[\mathcal{G}^{0}]_{\alpha_i\alpha_i}$. We then obtain all other diagonal blocks of \textit{forward connected} Green's function. This forward recursion halts when we have obtained $[\mathcal{G}^{F}]_{\alpha_{N+1}\alpha_{N+1}}$, which is for the rightmost sector. Since there are no further sectors, it can be shown that  $[\mathcal{G}^{F}]_{\alpha_{\mathcal{N}+1}\alpha_{\mathcal{N}+1}}= [\mathcal{G^N}]_{\alpha_{\mathcal{N}+1}\alpha_{\mathcal{N}+1}}$, the retarded Green's function of the $\alpha_{\mathcal{N}+1}$ sector.

	(b) From $[\mathcal{G}]_{\alpha_{\mathcal{N}+1}\alpha_{\mathcal{N}+1}}$, all other diagonal blocks of the retarded Green's function for all $\alpha_i$ sectors can be obtained by a backward recursion equation,
	\begin{eqnarray}
		[\mathcal{G^N}]_{\alpha_{i-1}\alpha_{i-1}}=[\mathcal{G}^{F}]_{\alpha_{i-1}\alpha_{i-1}}~~~~~~~~~~~~~~~~~~~~~~~~~~~~~~~~\nonumber\\
		\times(I+[\uptau]_{\alpha_{i-1}\alpha_i}[\mathcal{G^N}]_{\alpha_{i}\alpha_{i}}[\uptau]_{\alpha_i\alpha_{i-1}}[\mathcal{G}^{F}]_{\alpha_{i-1}\alpha_{i-1}})
		\label{eqn3}
	\end{eqnarray}
	
	(c) From the diagonal blocks of the retarded Green's function, we can calculate all off-diagonal blocks by the recursive relation,
	\begin{align}
		[\mathcal{G^N}]_{\alpha_i\alpha_j}|_{\alpha_i<\alpha_j}=-[\mathcal{G}^{F}]_{\alpha_i\alpha_i}[\uptau]_{\alpha_i\alpha_{i+1}}[\mathcal{G^N}]_{\alpha_{i+1}\alpha_j}
		\label{eqn4}
	\end{align}
	
	To summarize, the run starts with identifying $[\mathcal{G^{F}}]_{\alpha_{1},\alpha_{1}}$ to $[\mathcal{G}^{0}]_{\alpha_{1},\alpha_{1}}$ and calculating all $[\mathcal{G^F}]_{\alpha_{i}\alpha_{i}}$ by Eq.\ref{eqn2}. Then identifying $[\mathcal{G^F}]_{\alpha_{\mathcal{N}+1}\alpha_{\mathcal{N}+1}}$ with $[\mathcal{G^N}]_{\alpha_{\mathcal{N}+1}\alpha_{\mathcal{N}+1}}$, Eq.\ref{eqn3} computes all diagonal sectors of $[\mathcal{G}^\mathcal{N}]_{{\alpha_i}\alpha_i}$. Eq.\ref{eqn4} generates off-diagonal blocks $[\mathcal{G^N}]_{{\alpha_i}\alpha_j}(=[\mathcal{G^N}]_{{\alpha_j}\alpha_i}$).

	We make some important remarks: 
	(i) Matrix inversion is needed \textit{only for the forward recursion}. We use the standard linear solver ZGESV in the Intel MKL library for this computation. 
	(ii) We need only two matrices ($[\mathcal{G}^0]^{-1}_{\alpha_{i}\alpha_{i}}$, $[\mathcal{G}^{F}]_{\alpha_{i-1},\alpha_{i-1}}$) for computing $[\mathcal{G}^F]_{\alpha_{i}\alpha_{i}}$ during the forward recursion. Similarly, ($[\mathcal{G^N}]_{\alpha_{i}\alpha_{i}}$, $[\mathcal{G}^{F}]_{\alpha_{i-1},\alpha_{i-1}}$) are needed to compute $[\mathcal{G^N}]_{\alpha_{i-1}\alpha_{i-1}}$ during the backward recursion. Finally only two matrices are required for computing off-diagonal blocks as seen in Eq.\ref{eqn4}.
	The matrix multiplications to obtain terms like $[\uptau]_{\alpha_i\alpha_{i-1}}[\mathcal{G}^{F}]_{\alpha_{i-1}\alpha_{i-1}}[\uptau]_{\alpha_{i-1}\alpha_i}$ in the forward recursion are calculated in a manner that the connection matrices and the relevant Green's function matrices are not allocated in the memory simultaneously. The same approach is used for the backward recursion and the off-diagonal block Green's function calculations. This approach guarantees that at every step of the algorithm, \textit{two matrices are needed to be stored in the RAM.} 
	
We briefly discuss the model and sparseness independence of the F-RGF technique. The sectors $\alpha_i$ are labeled by fermion-occupations of the two halves of the lattice in Fig. 1(a), and the connection matrices facilitate fermion number fluctuations across the geometric divide. In this way of organizing the Hilbert space,  the model-specific origin of the source of fermion occupations across the geometric divide becomes irrelevant. The Fock-space structure remains invariant to fermion hopping across the divide due to long-range hopping, periodic boundary, or open boundary conditions. These model-dependent sources of hopping can make the Hamiltonian a dense matrix; however, the Fock-space structure, the dimensions of the sectors, and connection matrices depend solely on the lattice size and the total number of fermions. For example, for all-to-all hopping, almost all elements of the sector-Hamiltonians and the connection matrices will be non-zero; however, their \textit{sizes} remain fixed for a given filling and lattice size.
We also note that non-local fermion density-density dependent (long-range Coulomb interaction) is \textit{local} in the Hilbert space and hence would contribute only to the diagonal term of the matrix representation. Thus, such interaction terms are contained within the sector Hamiltonian $H(\alpha_i)$. More generally, problems where the fermion interaction term maps the many-fermion basis to itself (apart from a multiplicative factor) respect the block-tridiagonal form of the Hamiltonian. 

Finally, it is tempting to divide the real-space lattice into more than two partitions to reduce the sector dimensions. However, in these cases, the Fock-space sectors get coupled by more extended range hopping matrices, and the block-tridiagonal structure is lost. We would also like to point out that bi-partitioning is also employed in the Density-Matrix Renormalisation Group (DMRG)~\cite{dmrg-rmp}. However, the bi-partition  in F-RGF is used merely to group sets of basis states so that the Hamiltonian becomes tridiagonal. There is no approximation made in F-RGF, unlike DMRG only where maximally entangled states across the bipartition are retained to limit the exponential growth of the Hilbert space. Also, as discussed above, long-range hopping (which can generate a two-dimensional lattice) adds no memory overhead or error in F-RGF. 
	
	\begin{figure*}[t]
		\centering{
			\includegraphics[width=17cm, height=7cm, clip=true]{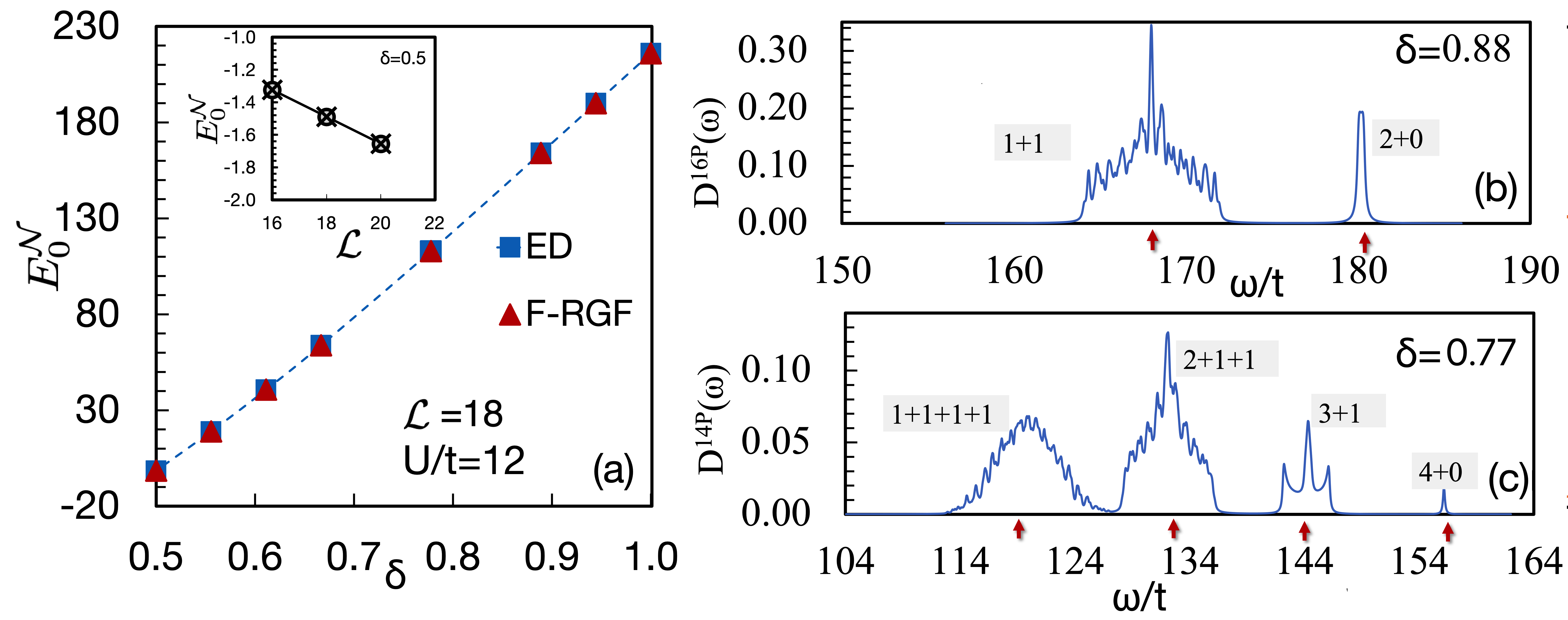}}
		\caption{\textbf{Ground-state energy benchmark \& many-fermion density of states for spinless interacting fermion model:} (a) shows the filling ($\delta$) dependence of ground-state energy ($E^{\mathcal{N}}_{0}$) for $\mathcal{L}=18$ sites periodic chain for $U=12.0t$. Inset shows (circles) $E^{\mathcal{N}}_{0}$ for different system sizes at half-filling. The crosses in the inset denote $E^{\mathcal{N}}_{0}$ from Bethe ansatz. $\mathcal{D}^{16P}(\omega)$ in (b) and $\mathcal{D}^{14P}(\omega)$ in (c) shows the 16-fermion and 14-fermion density of states for 18 site system for $U/t=12$. These correspond to fermion filling of $\mathcal{N}/\mathcal{L}$ of 0.88 and 0.77 respectively in (b) and  (c). Calculations are performed on Intel E5-2683v4 2.10GHz processors requiring $\sim 40$ mins ($\sim 13$ hours) per frequency point for $\mathcal{L}=18~(20)$ at half-filling.
		}
		\vspace{-0.0cm}
		\label{f-2}
	\end{figure*}
	\textit{iii. Computational advantage of F-RGF:} 
	
\textit{	a. Memory :} F-RGF replaces ($^\mathcal{L}C_\mathcal{N}$)-dimensional matrix inversion by $(^{\mathcal{L}/2}C_{i-1}\times^{\mathcal{L}/2}C_{\mathcal{N}-{i+1}})$-dimensional matrix inversions where $i\in (1,\mathcal{N}+1)$, regardless of the model or matrix-sparseness. Note that \textit{two matrices} are required in memory at every recursion step, as seen from Eq.\ref{eqn2} to Eq.\ref{eqn4}. Since the middle ($\alpha_{\mathcal{N}/2+1}$)-sector attains the largest dimension ($^{\mathcal{L}/2}C_{\mathcal{L}/4}\times ^{\mathcal{L}/2}C_{\mathcal{L}/4}$) for half-filling, it defines the RAM upper bound. $\mathcal{L}$ dependence of its dimension compared to that of the Hilbert space results in the memory efficiency of F-RGF as seen in Fig.~\ref{f-1} (e). 
	Fig.~\ref{f-1} (f) shows comparison of the exact analytical $(16/\pi\mathcal{L})$ scaling of the ratio of maximum F-RGF to DI RAM with numerical data. The scaling should allow access to $\mathcal{L}=22$ at half-filling and larger sizes at other fillings with present-day resources.

\textit{	b. Computaton time :} 
Many-fermion Green's function per frequency point in F-RGF distributes the total computation cost measured in terms of number floating point operations differently than in ED or direct inversion (per frequency point). 
The exact theoretical cost of computation in F-RGF can be expressed as $\sum_{i}\left([D(\alpha_i)]^3+6[D(\alpha_i)]^2+2[D(\alpha_i)]\right)$, where $[D(\alpha_i)]$ is the matrix dimension of sector $\alpha_i$ and $i$ runs over all sectors. The second term contain two matrix multiplications for computing $[\uptau]_{\alpha_i\alpha_{i-1}}[{\mathcal{G}}^{F}]_{\alpha_{i-1}\alpha_{i-1}}[\uptau]_{\alpha_{i-1}\alpha_i}$ in the forward recursion and three matrix multiplications for calculating $[\uptau]_{\alpha_{i-1}\alpha_i}[\mathcal{G^{N}}]_{\alpha_{i}\alpha_{i}}[\uptau]_{\alpha_i\alpha_{i-1}}[\mathcal{G}^{F}]_{\alpha_{i-1}\alpha_{i-1}}$ during the backward recursion. Finally, two matrix additions are required for computing the matrix summations in Eq. 2 and  Eq. 3. In contrast, the computation cost for full diagonalization (ED) or direct inversion (DI)  scales as $ (^\mathcal{L}C_\mathcal{N})^3$. We contrast the F-RGF and ED or DI computation cost for $\mathcal{L}/2$ spinless fermions on $\mathcal{L}$ sites or at half-filling.

The main panel in (g) (dashed line) shows this theoretical (dashed line) computation time ratio of F-RGF to ED or DI as a function of $\mathcal{L}$. The $\mathcal{L}$ dependence is better understood by plotting the inverse of this ratio (dashed line in inset). The inset shows a linear increase with $\mathcal{L}$. Consequently, the F-RGF compute time is suppressed as $\sim O(1/\mathcal{L})$ compared to ED or DI. It can be numerically verified that this $\sim O(1/\mathcal{L})$ scaling in the main panel is due to the dominant contribution of the first term (involving only inversions) for large $\mathcal{L}$ in the F-RGF expression of computation cost described above. In particular, for $\mathcal{L}\geq 16$ at half-filling, the most significant contribution comes from the inversion cost of the central sector $\alpha_{\mathcal{N}/2+1}$  (see Fig. 1(b)) and a few surrounding sectors. Since the $[D(\alpha_{\mathcal{N}/2+1})]=((^{\mathcal{L}/2}C_{\mathcal{L}/4})^2)$ at half-filling, the F-RGF to ED time ratio would be identical to the memory scaling if only contribution from the $\alpha_{\mathcal{N}/2+1}$ sector is considered. Due to contributions from other sectors of smaller dimensions, the $O(1/\mathcal{L}))$ scaling is exact only asymptotically at large $\mathcal{L}$, with a prefactor different from the memory scaling. The numerical data from code execution time (symbols) also follows similar linear behavior seen in the inset. Thus the time ratio of F-RGF over ED from the code execution time (symbols) also scales as $1/\mathcal{L}$ in the main panel (g). However, the execution time includes contributions from other code overheads, consequently giving a less significant advantage over the theoretical scaling.
	
	\section{Application}
	\label{sec:app}
	We consider spinless fermion model on periodic $\mathcal{L}$ site chain:
	\begin{align}
		H=-t\sum\limits_{\langle I,J\rangle }(c_{I}^{\dagger}
		c_{J}+h.c)+U\sum\limits_{I} n_{I} n_{I+1}
		\label{eqn5}
	\end{align}
	$c^\dagger_{I}$ ($c_{I}$) are fermion creation (annihilation) operators at site $I$; $t(\equiv 1)$ and $U$ are nn hopping and interaction respectively, $n_{I}=c^\dagger_{I}c_{I}$. 
	Here we focus on three specific quantities that can be computed from the resolvent, the many-fermion ground state energy,  the many-fermion density of states and few-fermion excitations in many-fermion systems.

\textit{(a) Ground state energy:} Since F-RGF is numerically exact, F-RGF $\mathcal{N}$-fermion ground-state energy ($E^\mathcal{N}_0$) is expected to match with exact diagonalization. Fig.~\ref{f-2}(a) shows typical data of $E^\mathcal{N}_0$ with filling $\delta(=\mathcal{N}/\mathcal{L})$ for $\mathcal{L}=18,~U/t=12$. In Section 2 we have discussed how $E^\mathcal{N}_0$ is determined in F-RGF from $[\mathcal{G^N}(\omega)]$. Inset demonstrates considerable $\mathcal{L}$ dependence of  $E^\mathcal{N}_0$ for $\delta=0.5$ up to $\mathcal{L}=20$. We also compare the F-RGF data shown in the inset with Bethe Ansatz (crosses) to demonstrate the accuracy of F-RGF further.

\textit{(b) Many-fermion density of states:} We now discuss many-fermion DOS results in  $\mathcal{D}^{16P}(\omega)$ Fig.~\ref{f-2} (b) and $\mathcal{D}^{14P}(\omega)$ in Fig.~\ref{f-2} (c) which are $(\mathcal{L}-2)$ and $(\mathcal{L}-4)$-fermion DOS respectively for $\mathcal{L}=18$.  $\mathcal{D}^{16P}(\omega)$ and $\mathcal{D}^{14P}(\omega)$ shows  16-fermion and 14-fermion density of states for $\mathcal{L}=18$ site system for $U/t=12$ . They are extracted from  $(\mathcal{L}-2)$ and  $(\mathcal{L}-4)$-fermion Green's functions for $\mathcal{L}=18$, as discussed in Sec. II.  In Fig.~\ref{f-2} (b), we have two features, one composed of the basis states where the two holes delocalise, avoiding being on nearest neighbor (nn) sites (1+1) and the other with the two holes on nn sites (2+0). The states contributing to the features can be identified by considering the zero hopping limit. The states belonging to the (1+1) hole configuration have a potential energy of $14U$, which is 4$U$ below the filled lattice energy of 18$U$. Similarly, the states belonging to the (2+0) hole configuration have a basis potential energy of  15$U$ or 3$U$ below the filled system energy of 18$U$. For $U/t=12$, these energy values define the centroids of the two features at $\omega/t=168$ and $\omega/t=180$ for the (1+1) and (2+0) features, respectively. Arrows mark the centroid locations in the figure. Hybridization effects in the presence of hopping provide widths to these features. The width of the lower energy feature is $\sim 8t$, and agrees with the convolution of two non-interacting spinless fermions. Similarly, in (c), we show the 14-fermion spectral function. It has four features labeled by (1+1+1+1), (2+1+1), (3+1), and (4+0). The states with no holes on nn sites have a dominant contribution to (1+1+1+1); two holes on nn sites and two non-adjacent holes make up the (2+1+1) feature. Three adjacent holes and one hole that is not nn to any other generate the (3+1) feature, and (4+0) consists of all four holes on nn sites. From left to right, the centroids of the features are at $\omega/t=8U$, 7$U$, 6$U$, and 5$U$ below the fully filled ground state energy $U\mathcal{L}$. Arrows mark the location of the centroids. The many-fermion DOS allows us to directly read off the ground state energy from the location of the lowest energy peak, as discussed in the previous section. In addition, as discussed above, the dominant many-body basis state contributing to any particular feature can be deciphered by calculating projected DOS for different classes of basis states. Such calculation can be carried out starting with two fermions, and gradually increasing the fermion number can help understand the evolution of few-body systems into that at a finite filling fraction. Moreover, the many-body DOS can allow the study of bound complexes for Hamiltonian with more complicated interactions as considered previously in literature \cite{few-mona-prl}. The F-RGF approach can also be used in presence of disorder to compute the many-fermion DOS for the study of many-body localization. We note that the fluctuation in the data in (b) and (c) arise out of finite size effects and can be reduced by going to larger system sizes.
	
	\begin{figure*}[t]
		\centering{
			\includegraphics[width=18cm, height=9.cm, clip=true]{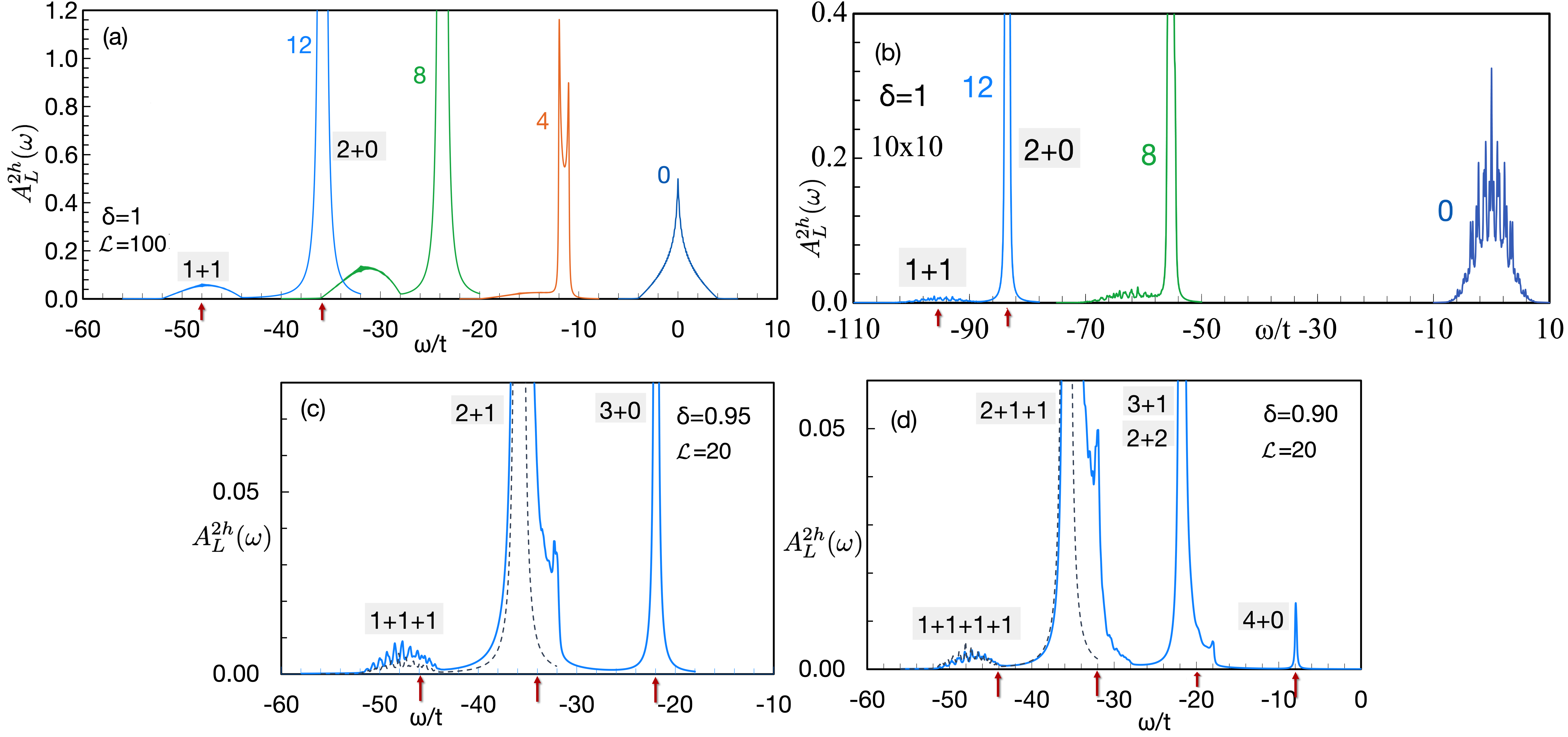}}
		\caption{\textbf{Two-hole excitation spectrum for spinless interacting fermion model:} (a)  and (b) show the evolution of $A^{2h}_{L}(\omega)$ with $U$, for $\delta=1$ in one and two dimensions.  (c) and (d) show $A^{2h}_{L}(\omega)$ for $\delta=$0.95 and $\delta=0.9$ respectively in one dimension. The dashed line in (c) and (d) is the $U=12t$ data for $\delta=1$, reproduced from (a) for comparison.  We note that the slight shift of the features away from the $\mathcal{N}-2$ basis state potential energy locations (arrows) with more ground state holes is due to increased hybridization effects among different classes of basis states. The various  class of basis states (labeled boxes) contributing to the features are provided in all panels.They are discussed in the text.
		}
		\vspace{-0.0cm}
		\label{f-3}
	\end{figure*}
\textit{(c) Few-hole excitation:} 
Few-body excitations play a vital role in the study of many-body physics. These include single particle or hole (photoemission) spectroscopy, two-particle local Auger electron spectroscopy (AES)\cite{aes-progress}, and dynamical charge susceptibilities. These are usually calculated by the Lanczos technique, which bypasses the need to calculate the full many-body DOS. For sparse Hamiltonians with short-range hopping and interaction strengths, Lanczos is highly advantageous with memory requirement scaling linearly in system size for ground states and few-body excitations. While the primary purpose of F-RGF is for calculating many-fermion resolvents, it is instructive to show how such quantities can be extracted within F-RGF formalism. 
As an example, we calculate the two-hole local density of states (L-DOS), which is measured in AES, for various situations. The standard many-body formula for few-body (fermion or hole) excitations can be easily derived in terms of the many-fermion density of states. We provide these in the Appendix. Here we show the results for local two-hole excitations in filled and partially-filled interacting fermion ground states in one and two dimensions. 

\underline{Two-hole spectrum in filled system:}  The two-hole L-DOS $A^{2h}_L(\omega)$, provides the excitation spectrum when two holes are added to adjacent site pair ($I, J$) in the $\mathcal{N}$-fermion ground state ($|\psi^{\mathcal{N}}_0 \rangle$) of $H$. As in Eq.\ref{eqn9} of the Appendix, expresses the L-DOS in terms of elements of  $\mathcal{N}$-fermion spectral function matrix $[\mathcal{D}^{\mathcal{N}}(\omega)]$ evaluated at $\omega=E^\mathcal{N}_0$ and $[\mathcal{D}^{\mathcal{N}-2}(\omega)] $ matrix elements. Importantly, the latter controls $\omega$ dependence of L-DOS. Fig.~\ref{f-3} (a) shows L-DOS for $U=0$, $4t$, $8t$ and $12t$ for ($\delta=1$) and $\mathcal{L}=100$ one-dimensional system. (b) shows L-DOS for two-dimensional system at the $U$ values indicated for $\delta=1$. All plots are shifted by respective $E^\mathcal{N}_0$ values. 

In (a), the $U=0$ L-DOS has a spread of about $8t$, expected for two non-interacting spinless holes in one-dimension \cite{few-mona-prl,few-anam,few-anam-2}. Increasing $U$ distorts and shifts the L-DOS to lower $\omega$. For $U\gtrapprox 4t$, a `split-off' resonance is seen at $\omega=-3U$ and a remnant feature of width $\sim 8t$ centered around $\omega=-4U$, agreeing with Cini-Sawatzky theory \cite{aes-sawat}. 
Since ($\mathcal{N}-2$)-fermion spectral function matrix controls the $\omega$-dependence of L-DOS, we analyze potential energies (PE) of $(\mathcal{N}-2)$-fermion basis states, i.e., the energy of basis states depending on hole positions for $t=0$, to identify where they contribute. For $\delta=1$, ($\mathcal{N}-2$)-fermion basis states have two holes. Basis states with non-adjacent holes collectively labeled as (1+1) have PE=$(\mathcal{L}-4)U$, and those with adjacent holes (2+0), have PE=$(\mathcal{L}-3)U$.  Accounting for the $E^\mathcal{N}_0(=\mathcal{L}U)$ shift for $\delta=1$, the features from (1+1) and (2+0) basis states occur at -4$U$ and -3$U$ respectively, marked (arrows) for $U=12t$ in (a). Fig.~\ref{f-3} (b) we show the two-hole spectral function in two-dimension for $U/t=0,8$ and 12 on $10\times 10$ lattice for $\delta=1$. For $U=0$, in 2D, the non-interacting two-hole L-DOS  bandwidth is expected to be close to 16$t$. We find that the $U=0$ spectral function conforms to this expectation. With increasing $U$, similar to the one-dimensional case, the spectral function splits into two features, one corresponding to the two holes delocalising independently (1+1) and the other with two holes delocalising while on nn sites. However, unlike in one dimension where the high energy split-off occurs beyond $U=4t$, the critical $U$ for creating a two-hole resonance is now found to be $8t$. The critical $U$ for the split-off resonance agrees with the expectation that the critical $U$ depends on the band minimum of the non-interacting many-fermion DOS \cite{aes-sawat}. Finally, the centroid locations of the (1+1) and (2+0)) features in two dimensions are at 8$U$ and 7$U$ below the ground state energy, respectively. Since the plots are shifted by $E^\mathcal{N}_0$, these features are located at $\omega/t=-96$ and $\omega/t=-84$ for $U=12t$ as marked in (b).

\underline{Two-hole spectrum in partially-filled system:} 
(c) and (d) show L-DOS for $U=12t$ and $\mathcal{L}=20$ one-dimensional lattice for partial fillings  $\delta=0.95$ and 0.9 with respective ground states containing one and two holes. These plots are also shifted by respective $E^\mathcal{N}_0$ values. New spectral features arising from partial filling can be identified as done above. In (c), $(\mathcal{N}-2)$-fermion basis states three-holes with (1+1+1), (2+1) and (3+0) hole configurations. They have $(\mathcal{L}-6)U$, $(\mathcal{L}-5)U$ and $(\mathcal{L}-4)U$ for PE. With a $E^{\mathcal{N}}_0=(\mathcal{L}-2)U-2t$ shift, the feature arising from these sets of basis states occurs at  $\omega=-4U+2t$, $-3U-2t$ and $-2U-2t$.
In (d), the $(\mathcal{N}-2)$-fermion basis states hole configurations are (1+1+1+1), (2+1+1), (2+2), (3+1) and (4+0) with PE= $(\mathcal{L}-8)U$, $(\mathcal{L}-7)U$, $(\mathcal{L}-6)U$, $(\mathcal{L}-6)U$ and $(\mathcal{L}-5)U$ respectively. Subtracting them from $E_0^\mathcal{N}$ defines the feature positions and allows identification of the basis states contributing to them. Computing partial trace of Eq.\ref{eqn9} over different classes of basis states can provide detailed line shapes, and hybridization between classes of basis states for all $\delta$ and $U$ but is not pursued in this proof-of-principle demonstration.

	\section{Summary}
We have presented a scheme that reorganizes the Hilbert space into a structured lattice in the Fock space allowing the use of the well-known recursive Green's function method. Due to the resulting block-tridiagonal representation of the Hamiltonian, only specific parts are needed in memory during the recursion steps. This feature crucially leads to a $O(1/\mathcal{L})$ suppression of the exponential growth of the Hilbert space with system size. We have demonstrated the scheme by explicit example calculations in one and two in two dimensions. We have also argued that the F-RGF scheme does not require any assumption of Hamiltonian symmetry or boundary conditions. We have demonstrated how the many-fermion DOS from the F-RGF scheme can be used to calculate several quantities of interest for many-body physics in one and two dimensions. While traditionally, few-body excitations in many-body ground states have dominated the exploration of many-body physics, several fundamental problems require the full many-fermion density of states. The many-fermion density of states can allow access to thermodynamic properties. Also, the many-fermion DOS can be used to study many-body localization, the onset of chaos in quantum systems \cite{few-chaos}, bound complexes in cold atomic  \cite{2h-bs-nat}, and molecular systems  \cite{q-chemistry}. The scheme can be extended to spin full-fermions and spin systems such as the XXZ model. Implementing symmetries in the F-RGF scheme will allow access to larger system sizes. 

\appendix*
 \setcounter{equation}{0}  
\begin{center}
	\textbf{APPENDIX}
\end{center}

\textit{Formula for two-hole spectral function:} For two holes created on a pair of lattice sites ($I, J$) in a $\mathcal{N}$-fermion ground-state $|\psi_0^\mathcal{N}\rangle$ of $H$ and subsequently destroyed at a later time from the same site-pair, the two-hole retarded Green's function, in the frequency domain is defined as:
	\begin{align}
		G^{{2h}}_{IJ;IJ}(\omega)=\sum_{j_\mathcal{N},j_\mathcal{N}^\prime}\langle\psi_0^\mathcal{N}|j_\mathcal{N}\rangle\langle j_\mathcal{N}^\prime|\psi_0^\mathcal{N}\rangle\nonumber\\
		\times \langle j_\mathcal{N}| c^\dagger_{I}c^\dagger_{J} ((\omega+ i\eta)I-H)^{-1}c_{I}c_{J}|j_\mathcal{N}^\prime\rangle
		\label{eqn6}
	\end{align}
	In the above we have inserted a complete set of $\mathcal{N}$-fermion real-space basis sets $\{|j_\mathcal{N}\rangle\}$ and $\{|j^{\prime}_\mathcal{N}\rangle\}$. The two-hole spectral function is obtained from the imaginary part of the above expression. We first provide a way to obtain the pre-factors $\langle \psi_0^\mathcal{N}|j_\mathcal{N}\rangle\langle j^\prime_\mathcal{N}|\psi_0^\mathcal{N}\rangle$ in Eq.\ref{eqn6}, from the $\mathcal{N}$-fermion Green's function.
	The imaginary part of $\mathcal{N}$-fermion Green's function in the Lehmann representation in Eq.\ref{eqn1} can be expressed as: 
	
	\begin{align}
		-\frac{1}{\pi}\Im{\mathcal{G^{\mathcal{N}}}_{j_{\mathcal{N}};j^\prime_{\mathcal{N}}}(\omega)}=\sum_{\lambda^{\mathcal{N}}}\langle j_{\mathcal{N}}|\lambda^{\mathcal{N}}\rangle \langle\lambda^{\mathcal{N}}|j^\prime_{\mathcal{N}}\rangle\nonumber\\ \times\delta(\omega-E^{\mathcal{N}}_{\lambda}) ~~~~~~~~~~~~~~~
		\label{eqn7}
	\end{align}
In practise we can extract $\langle \psi_0^\mathcal{N}|j_\mathcal{N}\rangle\langle j^\prime_\mathcal{N}|\psi_0^\mathcal{N}\rangle$  in two equivalent ways. We substitute $\omega=E_0^\mathcal{N}$ in the left hand side of \ref{eqn7} and scale the result by $\eta\pi$. This is because the maximum value of the Lorentzian representation of  $\delta(\omega-E_0^\mathcal{N})$ is $1/\eta\pi$ at $\omega=E_0^\mathcal{N}$. Here $\eta$ is the same broadening factor used for computing the many-fermion Green's function. Another equivalent way is to consider a small window around the $\omega=E_0^\mathcal{N}$ and integrate the left hand side of \ref{eqn7} over this window, and use the area of the Lorenztian calculated separately over the same window as a normalization. We have numerically verified that both approaches yield identical results.
	
 Here we make the standard assumption of a symmetric Green's function matrix , \textit{i.e.}, $\mathcal{G}^{\mathcal{N}}_{j_{\mathcal{N}};j_{\mathcal{N}}^\prime}(\omega)=\mathcal{G}^{\mathcal{N}}_{j_{\mathcal{N}}^\prime;j_{\mathcal{N}}}(\omega)$. This property guarantees that $[D^{\mathcal{N}}(\omega)]$ is a real quantity. Under this condition, the imaginary part of  
	$G^{{2h}}_{IJ;IJ}(\omega)$ in Eq.\ref{eqn6}, is given by the imaginary part of $\langle j_\mathcal{N}| c^\dagger_{I}c^\dagger_{J} ((\omega+ i\eta)I-H)^{-1}c_{I}c_{J}|j_\mathcal{N}^\prime\rangle$. This imaginary part is just the  matrix elements of $(\mathcal{N}-2)$-fermion spectral function matrix $[\mathcal{D}^{\mathcal{N}-2}(\omega)]$. $A^{2h}_L(IJ;IJ\omega)\equiv -1/\pi Im\{G^{{2h}}_{IJ;IJ}(\omega)\}$, the two hole L-DOS can then be expressed as: 
	\begin{align}
		A^{2h}_L(IJ;IJ,\omega)=\sum_{j_{_{\mathcal{N}}},j^\prime_{_{\mathcal{N}}}
		}\mathcal{D}^{\mathcal{N}}_{j_{_{\mathcal{N}}},j^\prime_{_{\mathcal{N}}}}(E^
		{\mathcal{N}}_0)\mathcal{D}^{\mathcal{N}-2}_{j_{_
				{\mathcal{N}}}(IJ)^-;j^\prime_{_{\mathcal{N}}}(IJ)^-}(\omega)
		\label{eqn9}
	\end{align}
	
	For the periodic lattice with translation invariance studied here, we have suppressed the ($I, J$) labels in the definition of L-DOS, simply as $A^{2h}_L(\omega)$ in what follows for brevity of notation.  
	From Eq.~\ref{eqn9} we first notice that calculation of real space two-hole  spectral function $A^{2h}_L(\omega)$, involves elements of the $\mathcal{N}$-fermion spectral function matrix, $\mathcal{D}^{\mathcal{N}}
	_{j_{_{\mathcal{N}}},j^\prime_{_{\mathcal{N}}}}(\omega)$ evaluated at $\omega=E^{\mathcal{N}}_0$ or at the many-fermion ground-state energy. Different elements of $\mathcal{D}^{\mathcal{N}}_{j_{_{\mathcal{N}}},j^\prime_{_{\mathcal{N}}}}(\omega)$ are extracted from the many fermion Green's function $\mathcal{G}^{\mathcal{N}}_{j_{_{\mathcal{N}}};j^\prime_{_{\mathcal{N}}}}(\omega)=\langle j_{_{\mathcal{N}}}|\mathcal{\hat{G}(\omega)} |j^\prime_{_{\mathcal{N}}}\rangle$, evaluated between the $\mathcal{N}$-fermion basis elements, $(|j_{_{\mathcal{N}}}\rangle)^\dagger$ and $|j_{_{\mathcal{N}}}^\prime\rangle$ at $\omega=E^{\mathcal{N}}_0$. 
	The $(\mathcal{N}-2)$-fermion spectral function matrix elements required in Eq.~\ref{eqn1}, are similarly extracted from $[\mathcal{G}^{\mathcal{N}-2}(\omega)]$.
	The $j_{_{\mathcal{N}}},j^\prime_{_{\mathcal{N}}}$ indices run over all the $\mathcal{N}$-fermion basis-states, while the relation  $|j^\prime_{_{\mathcal{N}}}(IJ)^-\rangle\equiv c_Ic_J|j_{_{\mathcal{N}}}^\prime\rangle$ defines the $(\mathcal{N}-2)$-fermion basis indices in Eq.~\ref{eqn9}. The unprimed indices refer to corresponding conjugate states and are defined in an analogous fashion. Finally, we note that two-particle excitations in partially-filled bands can be computed from $\mathcal{N}$ and  $(\mathcal{N}+2)$-fermion spectral function matrices. Similarly,  one (particle/hole) photoemission excitations can also be computed from $\mathcal{N}$ and  ($\mathcal{N}+1/\mathcal{N}-1$) -fermion spectral function matrices.
	
\section*{Acknowledgements}
	
	We acknowledge the use of NOETHER and VIRGO clusters at NISER.
	We acknowledge funding from the Department of Atomic Energy, India under Project No. 12-R\&D-NIS-5.00-0100.
		
	\bibliography{bibliography.bib}

\end{document}